\documentclass[aps,pra,superscriptaddress,twocolumn,5p]{revtex4-2} %mudar para twocolumn no paper

\usepackage{amsmath}
\usepackage{amssymb}
\usepackage{amsthm}
\usepackage{amsfonts}
\usepackage{bbm}
\usepackage{xcolor}
\usepackage{graphicx}[floatfix]
\usepackage{hyperref}
\usepackage[utf8]{inputenc}
\usepackage[T1]{fontenc}
\usepackage{physics}
\usepackage{times}
\usepackage{cancel}
\usepackage{wasysym}
\usepackage{comment}
\usepackage{ulem}
\usepackage{mathtools}
\usepackage{float}

\hypersetup{
	colorlinks=true,linkcolor=blue,citecolor=blue,
	filecolor=blue,urlcolor=blue,breaklinks=true
}
\newcommand{\largep}[1]{\left(#1\right)}
\newcommand{\orcid}[1]{\href{https://orcid.org/#1}{\includegraphics[width=7pt]{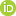}}}

\theoremstyle{definition}
\newtheorem{definition}{Definition}[section]

\begin{document}
	
\title{$\mathbf{k}$-uniform complete hypergraph states stabilizers in terms of local operators}

\author{Gabriel M. Arantes\orcid{0009-0004-7382-2635}}
\email{gabriel.moniz@usp.br}
\affiliation{
	Department of Mathematical Physics, Institute of Physics, University of São Paulo, Rua do Matão 1371, São Paulo 05508-090, São Paulo, Brazil
}

\author{Vinícius Salem\orcid{0000-0002-1768-8783}}
\email{vinicius.salem@uva.es}
\affiliation{
	Departamento de F\'{i}sica Te\'{o}rica, At\'{o}mica y \'{O}ptica, Universidad de Valladolid,
	47011 Valladolid, Spain
}

\author{Danilo Cius\orcid{0000-0002-4177-1237}}
\email{danilocius@gmail.com}
\affiliation{
	Department of Mathematical Physics, Institute of Physics, University of São Paulo, Rua do Matão 1371, São Paulo 05508-090, São Paulo, Brazil
}

\author{Barbara Amaral\orcid{0000-0003-1187-3643}}
\email{bamaral@if.usp.br}
\affiliation{
	Department of Mathematical Physics, Institute of Physics, University of São Paulo, Rua do Matão 1371, São Paulo 05508-090, São Paulo, Brazil
}
\affiliation{Department of Chemistry and Centre for Quantum Information and Quantum Control,
University of Toronto, 80 Saint George St., Toronto, Ontario, M5S 3H6, Canada}

\date{\today}
	
\begin{abstract}

In this work, we present a novel method to express the stabilizer of a $\mathbf{k}$-uniform complete hypergraph state as a linear combination of local operators. Quantum hypergraph states generalize graph states and exhibit properties that are not shared by their graph counterparts—most notably, their stabilizers are intrinsically nonlocal, as hyperedges can involve arbitrary subsets of vertices. Our formulation provides an explicit description of the stabilizers for $\mathbf{k}$-uniform complete hypergraphs and may offer new insights for exploring these states within the stabilizer formalism. In particular, this approach could facilitate the construction of new Bell inequalities or find applications in quantum error correction.
\end{abstract}
	
\maketitle
%\tableofcontents
\pagebreak

% \danilo{
% Comandos para inserções:
% \begin{enumerate}
%     \item \gabriel{Gabriel}: \textbackslash gabriel\{ texto \} 
%     \item \vinicius{Vinicius}: \textbackslash vinicius\{ texto \}
%     \item \danilo{Danilo}: \textbackslash danilo\{ texto \}
%     \item \barbara{Barbara}: \textbackslash barbara\{ texto \}
% \end{enumerate}

%  \revision{revisão}: \textbackslash revision\{ texto \}
% }

\section{Introduction}
\label{sec:intro}

Quantum nonlocality, one of the main phenomena discovered in physics over the last six decades, is associated with the fact that quantum systems can be correlated in such a way that it is impossible to reproduce in classical systems, as is widely known in the form of Bell's theorem \cite{Bell_1964}. Since then, this feature has been extensively studied and applied as a resource for quantum communication and quantum technologies \cite{Zapatero_2023}. 
There is a strict relationship between nonlocality and self-testing, a secure method of verifying both the local measurements performed by a quantum device and its final produced state, only by analyzing the resulting correlations \cite{Mayers_2003, Supic_2020}, allowing one to understand the structure of the set of quantum correlations of a given system. Such correlations allow one to certify that the device is working properly without the need to access its internal structure; therefore, it is usually referred to as device-independent (DI) or box certification. 

Hypergraph states (HS) are a natural generalization of graph states (GS), which constitute a suitable class of pure quantum states for studying quantum entanglement and nonlocality \cite{rossi2013quantum,JPA.47.335303.2014, Noller_2023} and recently implemented experimentally for quantum computation \cite{Vigliar_2021,SR.9.1.2019, huang2024demonstration}. Nonlocal properties can be analyzed for GS using Bell inequalities derived from stabilizer operators, as demonstrated in \cite{Guhne_2005}. Furthermore, scalable Bell inequalities have been constructed for GS, as shown in \cite{Baccari_2020}. However, these methods cannot be directly applied to HS due to the inherently nonlocal nature of their stabilizing operators. Still, local correlations derived from nonlocal stabilizers can reveal violations of local realism. For instance, in Ref. \cite{Gachechiladze_2016}, the authors employed Hardy's argument to demonstrate the nonlocality of HS. %A particularly interesting subclass, known as symmetric HS, possesses local Pauli stabilizer operators. This property makes them strong candidates for developing new Bell inequalities, as discussed in \cite{Lyons_2017, Noller_2023}. 

Also, HS possesses unique characteristics not shared by GS, such as computational universality for MBQC under only
Pauli measurements \cite{SR.9.1.2019}. Despite being initially pure qubit states, generalizations for mixed HS and qudit HS are also possible \cite{PhysRevA.109.012416,PhysRevA.95.052340}, and despite being characterized by its stabilizers, the nonstabilizerness of HS might be useful for investigating quantum state complexity and 
fault-tolerant \cite{chen2024magic}.

Symmetric HS \cite{IJTP.58.1.2019,JPA.48.95301.2015} are particularly interesting due to their direct parallel to generalized GHZ states. In fact, this study aims to express the non-local stabilizer of a $k-$uniform complete HS as a linear combination of local operators with the goal that this enables a new Bell inequality.
Finally, it has been shown that HS have a large potential
in Gaussian quantum information, which may be quite appealing concerning the possibility of the efficient verification of quantum
states in this scenario \cite{PRL.126.240503.2021,PRA.100.062301.2019}.

The first construction of Bell inequalities using GS through their stabilizer formalism proved to be very simple and effective \cite{Guhne_2005}. However, the fact that it scales exponentially with the number of parties made it a challenge for experimental implementation.  
Generalizing for HS poses an even greater difficulty because the notion of nonlocality can only be established by making local measurements; however, the HS is usually expressed using non-local operators. This paper presents a method to express HS stabilizers in terms of a linear combination of local operators, with the hope that this approach may lead to a Bell inequality or be applied in other contexts where the stabilizer formalism is useful, such as error correction \cite{PhysRevLett.95.230504}. 

We obtain an exact closed-form expression for the coefficients \( C_m \) (Eq.~\eqref{eq final}) and identify a sign structure that hinders the direct application of the Bell-inequality construction in \cite{Baccari_2020} for \( k > 2 \). Consequently, the formulation of Bell inequalities in this case requires an alternative approach.

\section{Quantum Hypergraph States}
\label{sec:hypergraphs}

Before presenting the physical correspondence between hypergraphs and quantum states, a concise mathematical definition of a hypergraph is necessary. A hypergraph is a generalization of a graph as follows \cite{Book.2001.Berge,graphsandhypergraphs,voloshin2009introduction}:
\begin{definition}
  A hypergraph is a pair of sets $(V, E)$, where $V = \{v_1, \dots, v_N\}$ denotes the set of vertices and $E$ is the set of hyperedges, with each element $e \in E$ satisfying $e \subset V$ and $|e| \geq 2$ ($|X|$ denotes the number of elements of a finite set $X$).
\end{definition}

\begin{definition}
  A $k$-uniform hypergraph $(V, E)$ is a hypergraph such that $|e| = k$ for all $e \in E$.
\end{definition}

The uniformity condition imposes an additional structure on a hypergraph. Another important concept that will be useful later is defined as follows.

\begin{definition}
  A complete $k$-uniform hypergraph $(V, E)$ is a $k$-uniform hypergraph, in which every subset of $V$ with $k$ elements is a hyperedge.
\end{definition}

\begin{figure}[h]
    \centering
    \includegraphics[width=0.8\linewidth]{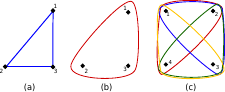}
    \caption{Example of a graph and its generalizations as hypergraphs. (a) A graph with 3 vertices demands at least 3 edges to be complete. (b) Its generalization as a hypergraph, however, demands only one hyperedge to be complete. (c) A 3-uniform complete hypergraph with 4 vertices as a generalization of the GHZ state; the state represented by this hypergraph corresponds to the GHZ state.}
    \label{fig:fig1}
\end{figure}

From this point, we consider $V = \{1, \dots, N\}$. A hypergraph can be associated with a quantum state through the following construction:
\begin{equation}
    |H\rangle = \prod_{e \in E} CZ_e \, \ket{+}^{\otimes N},
\end{equation}
where $CZ_e$ is a generalized controlled-phase gate acting on the qubits indexed by the elements of $e$
\begin{equation}  
CZ_{e} = \mathbb{I}_{e_1}\mathbb{I}_{e_2}\cdots\mathbb{I}_{e_n} - 2 \ketbra{1_{e_1}\cdots 1_{e_n}}  
\end{equation} the resulting state is referred to as a \textit{HS}.

A graph corresponds to the special case, in which $|e| = 2$ for all $e \in E$. In contrast, a general hypergraph allows hyperedges of arbitrary sizes ranging from $2$ to $|V| = N$. 

The concept of uniformity can be further generalized by allowing multiple uniformities to be present simultaneously. 
\begin{definition}
A complete $\mathbf{k}$-uniform hypergraph $(V, E)$ is defined for a set $\mathbf{k} = (k_1, \dots, k_p)$ of distinct $k_i$, 
where the edge set $E$ consists of all subsets of $V$ with cardinalities $k_i$ for some $i$. 
\end{definition}
This construction includes, as special cases, both the complete $k$-uniform and the fully connected hypergraph. 
Such symmetric complete $\mathbf{k}$-uniform HS have been investigated in 
\cite{Symmetric-Hypergraph, JPA.50.245303.2017}, where a local stabilizer was found by applying products of the non-local stabilizer \eqref{stabilizer} and establishing a non-locality test using a Mermin inequality.  
% \revision{A symmetric complete k-uniform HS is \cite{JPA.50.245303.2017} EQUATIONS 4, 3, 5 PAPER Symmetric HS (guhne) } 
% This motivates the following definition:
\subsection{The stabilizer formalism}

Similar to GS, HS admit an alternative characterization through stabilizer formalism. However, a crucial distinction emerges: whereas GS have local stabilizers (tensor products of Pauli operators), HS are stabilized by \textit{nonlocal} operators. %These stabilizers incorporate nonlocal phase gates that act across multiple qubits. 

\begin{definition}
    Let $\mathcal{H}=(\mathbb{C}^2)^\otimes$ be the Hilbert space of $n$ qubits. The Pauli group for one qubit is generated by the following matrices:
\begin{equation}
     X = \left( \begin{array}{cc}
0 & 1 \\
1 & 0
\end{array} \right), \hspace{0.2cm}
Y=  \left( \begin{array}{cc}
0 & -i \\
i & 0
\end{array} \right),  \hspace{0.2cm}
Z = \left( \begin{array}{cc}
1 & 0 \\
0 & -1
\end{array} \right),
\end{equation}
\noindent where the n-fold tensor products of Pauli matrices and the identity $\{\mathbb{I},X,Y,Z\}^{\otimes N}$ form an orthogonal basis for the space of linear operators on $\mathcal{H}$ \cite{PhysRevResearch.2.043323}.
\end{definition}

Adding the phase factors $\{\pm 1 \pm i\}$ to the previous operators gives to us the Pauli group $P_n$ whose order is $4^{n+1}$.

The stabilizer formalism was introduced in \cite{gottesman1997stabilizer} for quantum error correction, and provides a setting for demonstrating that operations in quantum computation using the Clifford group are classically efficiently simulable \cite{PhysRevA.73.022334,PhysRevLett.91.147902}.
The nonlocal nature of these stabilizers distinguishes HS from conventional GS, as each $g_l$ involves multiqubit phase operations across all hyperedges incident to vertex $i$, as presented in the following.

\begin{definition}
    Given a hypergraph, let $N(l)=\{e'\ |\ e'\cup\{l\} \in E\ \text{and}\ e'\cap\{l\} = \emptyset \}$ be the set of neighbors of vertex $l$ for a complete $\mathbf{k}$-uniform hypergraph $N(l)
  = \left\{\, e' \subseteq V \setminus \{l\}
      \;\middle|\;
      \exists\, i \in \{1,\dots,p\} \text{ such that } |e'| = k_i - 1
    \right\}$. Its associated HS can also be defined by the unique state stabilized by the set of operators $\{g_l\}$, such that
\begin{equation}
    g_l = X_l \prod_{e'\in N(l)}CZ_{e'}
    \label{stabilizer}
\end{equation}
for $l\in V$, which means that the HS is the unique state that is an eigenstate with eigenvalue $+1$ of all $g_l$.
\end{definition}

Note the expression for the stabilizer \eqref{stabilizer} can be written as
\begin{equation*}
    g_l = \left( \prod_{e \in E} CZ_e \right) X_l \left( \prod_{e' \in E} CZ_{e'} \right).
\end{equation*} 
This establishes that the stabilizers form an Abelian group by satisfying the commutation relations arising from the Pauli group:
\begin{align}
    [g_i, g_j]  =   \left( \prod_{e \in E} CZ_e \right)[X_i,X_j] \left( \prod_{e' \in E} CZ_{e'} \right)=0
\end{align}
Furthermore, the stabilizers $g_i$ are traceless and form a maximal set of $N$ commuting observables, being the respective $HS$ the only state that has eigenvalue $+1$ for all of them:
\begin{equation}
    g_i \ket{H} = \ket{H}, \space \forall \space i.
\end{equation}

Through the definition of the stabilizing operators, we can present the stabilizer group:
\begin{equation}
\mathcal{S}=\bigg\{S_x \left| S_x = \prod_{i \in V}(g_i)^{x_i} \quad \text{and} \quad x \in \{0,1\}^N\right.\bigg\},
\end{equation}
\noindent which consists of all the distinct products of stabilizers (since $g_i^2 = I_i$). $\mathcal{S}$ is an Abelian group with $2^N$ elements that necessarily satisfies the condition:
\begin{equation}
S_x \ket{H} = \ket{H},
\end{equation}
\noindent that is, a HS is not only an eigenstate of each stabilizer $g_i$ with eigenvalue $+1$, but also of all possible combinations of these operators. It is now possible to write the HS projector in the form \cite{JPA.47.335303.2014}:
\begin{equation}
\ketbra{H} = \frac{1}{2^N} \sum_{x \in \{0,1\}^N} S_x = \prod_{i=1}^n \frac{\mathbb{I} + g_i}{2}
\end{equation}

$\mathcal{S}$ is a subgroup of $\mathcal{P}_n$ and is called the stabilizer group since it is commutative and does not contain $\mathbb{-I}$. Hence, it does not include a Pauli operator with phases $\pm i$, all the elements of this group have order 2 except the identity, meaning that $\mathcal{S}$ is isomorphic \cite{PhysRevResearch.2.043323}.

Another important aspect is how they are expressed on the computational basis. For $\tau \in \{0,1\}^N$, let $J(\tau) \coloneqq \{ i \mid \tau_i = 1 \}$. It follows that
\begin{equation}
\begin{split}
    e \subset J(\tau)  &\implies CZ_e|\tau\rangle = -|\tau\rangle,\\
    e \not\subset J(\tau) &\implies CZ_e|\tau\rangle = |\tau\rangle.
\end{split}
\end{equation}
Define \( n_E(\tau) = |\{ e \mid e \in E \ \text{and} \ e \subset J(\tau) \}| \), such that
\begin{equation}
\begin{split}
    |HG\rangle &= \prod_{e \in E} CZ_e \frac{1}{\sqrt{2^N}} \sum_{\tau \in \{0,1\}^N}|\tau\rangle \\
    &= \frac{1}{\sqrt{2^N}} \sum_{\tau \in \{0,1\}^N} \prod_{e \in E} CZ_e|\tau\rangle \\
    &= \frac{1}{\sqrt{2^N}} \sum_{\tau \in \{0,1\}^N} (-1)^{n_E(\tau)}|\tau\rangle.
\end{split}
\label{computational basis hypergraph}
\end{equation}

For example, consider the hypergraph \((b)\) in Fig.~\ref{fig:fig1}, which consists of three vertices \(V = \{1, 2, 3\}\) and a single hyperedge \(E = \{V\}\). In this case, the only configuration \(\tau\) that yields a nonzero value of \(n_E(\tau)\) is \((1,1,1)\), where \(n_E(\tau) = 1\). The corresponding HS is therefore
\begin{align*}
    \ket{H} = \frac{1}{\sqrt{8}}\big(&\ket{000} + \ket{001} + \ket{010} + \ket{100} \nonumber\\
    &+ \ket{011} + \ket{101} + \ket{110} - \ket{111}\big).
\end{align*}

For a complete $k$-uniform hypergraph, there exists an explicit expression for $n_E(\tau)$, which is denoted as $n_k(\tau)$. Since every subset $e \subset V$ with size $k$ is an element of $E$, it is sufficient to count how many distinct $k$-tuples of "ones" are present in $\tau$. This leads to the following expression: 
\begin{equation}
{n_k(\tau)=}{|J(\tau)| \choose k}
\label{n_k}
\end{equation}
considering that ${m \choose k}=0$ when $m<k$.

It is clear that for a complete $\mathbf{k}$-uniform HS, with $\mathbf{k} = (k_1, \dots, k_p)$, it generalize as
\begin{equation}
{n_{\mathbf{k}}(\tau)=}\sum_{i=1}^p{|J(\tau)| \choose k_i}.
\label{n_k}
\end{equation}

\section{Expansion of the stabilizers in terms of Local Operators}

From these definitions, one can express the controlled-$Z$ gate solely in terms of the local Pauli operator $Z$ and the identity. This, in turn, allows the stabilizer of a HS to be written in terms of the same operators together with the Pauli operator $X$. 

Although this approach applies to any hypergraph, deriving a general expression for the coefficients quickly becomes intractable. In this work, we focus on $k$-uniform complete hypergraphs because they provide a straightforward way to obtain exact expressions for the coefficients for any given number of vertices.
\subsection{Generalized CZ Expansion}
Let $v$ be a set $v={v_1,  v_2 ,...,v_n}$ of positive integers, where $v_i \neq v_j$ for $i \neq j$. Using the identity $\mathbb{I}-Z=2|1\rangle \langle1|$ it is possible to express

\begin{equation}
\begin{split}
    CZ_v =& \mathbb{I} - 2|1_{v_1}...1_{v_n} \rangle \langle1_{v_1}...1_{v_n}|\\
    =& \mathbb{I} - 2 \largep{\frac{\mathbb{I}_{v_1}-Z_{v_1}}{2}}... \largep{\frac{\mathbb{I}_{v_n}-Z_{v_n}}{2}}\\
    =& \mathbb{I} - \frac{1}{2^{|v|-1}}\bigotimes_{i\in v}(\mathbb{I}_i - Z_i)\\
    =&  \mathbb{I} - \frac{1}{2^{|v|-1}}\largep{\mathbb{I} + \sum_{m=1}^{|v|}(-1)^m\largep{\sum_{\substack{v' \subset v \\ |v'|=m}}\bigotimes_{i\in v'}Z_i}}\\
    =& \largep{1- \frac{1}{2^{|v|-1}}}\mathbb{I}+ \sum_{m=1}^{|v|}\frac{(-1)^{m+1}}{2^{|v|-1}}\largep{\sum_{\substack{v' \subset v \\ |v'|=m}}\bigotimes_{i\in v'}Z_i}
    \label{CZ expansion}
\end{split}
\end{equation}
leading to the desired expansion in terms of local operators. 

Now consider a general hypergraph with N vertices, for which the stabilizers of the associated HS are given by
\begin{equation*}
    g_l = X_l \bigotimes_{v \in N(l)} CZ_v .
\end{equation*}
To expand these stabilizers, observe that each term in the tensor product is of the form $\bigotimes_{i\in v'}Z_i  \bigotimes_{i\in v''}Z_i$;\ $(\bigotimes_{i\in v'}Z_i) \mathbb{I}$ or is the identity $\mathbb{I}$. However, only linear terms on which subspace $v_i$ will appear since $Z^2=\mathbb{I}$, so we can always write 
\begin{equation*}
    g_l=X_l \largep{C_0 \mathbb{I} + \sum_{v \subset V\backslash\{l\}} C_v \bigotimes_{i\in v}Z_i}
\end{equation*}
for some coefficient $C_v$ specified by the subset $v$.

This means that any product of $CZ_v$ can be written as a linear combination of $\{\mathbb{I}, \bigotimes_{i\in v}Z_i\}$, and the choice edges set defines the coefficients and vice versa. Nevertheless, this expression on its own is of no use, but for any specific case, it is possible to obtain the coefficients. 

For a $\mathbf{k} -$uniform complete hypergraph, there is a symmetry between the exchange of any two vertices so the coefficient of each term, for a fixed $|v|$ should be the same; therefore, the expression simplifies to
\begin{equation}
    g_l=X_l \largep{C_0 \mathbb{I} +\sum_{m=1}^{|V|-1}C_m\sum_{\substack{v \subset V\backslash\{l\} \\ |v|=m}}\bigotimes_{i\in v}Z_i}
    \label{stabilizer eq}
\end{equation}
for certain coefficients that can be determined explicitly. 

Therefore, the stabilizer of a complete k-uniform hypergraph can be expressed as a linear combination of the stabilizers of all complete GS, whose vertices form subsets of the hypergraph.
In the following, we present an example for stabilizers of the symmetric HS with 4 vertices represented by $(c)$ in the figure (\ref{fig:fig1}), whose state can be considered as a generalized GHZ state.
\subsubsection{The hypergraph state $\ket{H^{4}_{3}}$}
Consider the case of a $3$-uniform hypergraph with $N = 4$ vertices. The corresponding hypergraph state can be written as
\begin{align*}
\ket{H^{4}_{3}} &= \prod_{e\in E} CZ_{e} \ket{+}^{\otimes 4} \\
&= CZ_{1,2,3}CZ_{1,2,4}CZ_{1,3,4}CZ_{2,3,4} \ket{+}^{\otimes 4},
\end{align*}
where the set of hyperedges is
\[
E = \{\{1,2,3\}, \{1,2,4\}, \{1,3,4\}, \{2,3,4\}\}.
\]

Following Eq.~\eqref{stabilizer}, and noting that for any vertex $l$ the neighborhood $N(l)$ consists of all pairs of vertices that do not include $l$, the stabilizer associated with vertex~$1$ is given by
\begin{align}
g_{1} &= X_{1}\prod_{e\in N(1)} CZ_{e} = X_1\, CZ_{2,3}CZ_{2,4}CZ_{3,4}.
\end{align}

Using the expansion formula \eqref{CZ expansion} for a two-qubit controlled-$Z$ gate,
\begin{equation*}
CZ_{i,j} = \tfrac{1}{2}\bigl(\mathbb{I} + Z_i + Z_j - Z_i Z_j\bigr),
\end{equation*}
we first obtain
\begin{equation*}
CZ_{2,3}CZ_{2,4} = \tfrac{1}{2}\bigl(\mathbb{I} + Z_2 + Z_3Z_4 - Z_2Z_3Z_4\bigr),
\end{equation*}
and, after including the third term,
\begin{equation*}
CZ_{2,3}CZ_{2,4}CZ_{3,4} = \tfrac{1}{2}\bigl(Z_2 + Z_3 + Z_4 - Z_2Z_3Z_4\bigr)
\end{equation*}
hence, the stabilizer generator associated with vertex~$1$ takes the form
\begin{equation}
g_1 = \tfrac{1}{2}X_1\bigl(Z_2 + Z_3 + Z_4 - Z_2Z_3Z_4\bigr).
\label{stabilizer H_3^4}
\end{equation}

Due to the symmetry of this hypergraph, obtaining the stabilizer for vertex $l$ is equivalent to exchanging the indices $1$ and $l$ in Eq.~\eqref{stabilizer H_3^4}.  
This provides an explicit example illustrating how Eq.~\eqref{stabilizer eq} is evaluated for a specific case.  
Moreover, we can verify that the coefficients in Eq.~\eqref{stabilizer H_3^4} coincide with those predicted by the general formula in Eq.~\eqref{eq final} for this configuration.

\subsection{Expansion Coefficients expression}

To find the analytic expression for the coefficients $C_m$ for a $\mathbf{k}-$uniform complete hypergraph with $\mathbf{k}=(k_1,...,k_p)$, consider the states of the form 
\begin{equation*}
    |\psi_m\rangle=|\underbrace{0...0}_\text{m} \underbrace{+...+}_\text{N-m}\ \rangle,
\end{equation*}
for $m=0,...,N$ and compute the expectation value of the stabilizer of the last vertex $g_N$—both in its original and expanded forms—on these states.

For the expanded form define $V_m=\{1,...,m\}$, it follows that 
\begin{equation}
{
\langle\ \underbrace{0...0}_\text{m} \underbrace{+...+}_\text{N-m-1}| \bigotimes_{i\in v}Z_i |\underbrace{0...0}_\text{m} \underbrace{+...+}_\text{N-m-1}\ \rangle }=
\begin{cases}
  0, & v \not\subseteq V_m \\
  1, & v \subseteq V_m 
\end{cases}
\end{equation}
hence taking the expectation value
\begin{equation}
    \begin{split}
        \langle g_N \rangle&=  \langle X_N \largep{C_0 \mathbb{I} +\sum_{n=1}^{|V|-1}C_n\sum_{\substack{v \subseteq V\backslash\{N\} \\ |v|=n}}\bigotimes_{i\in v}Z_i} \rangle\\  
        &= C_0 + \sum_{n=1}^{|V_m|}C_n\sum_{\substack{v \subseteq V_m \\ |v|=n}}\mathbb{I}\\
        &= C_0 + \sum_{n=1}^{m}C_n {m \choose n}.
        \end{split}
    \label{eq coef}
\end{equation}

While for the usual form, defining $\mathbf{k}-1 = (k_1-1,...,k_p-1)$ for convenience:
\begin{widetext}
\begin{equation}
    \begin{split}
        \langle g_N \rangle&=  \langle X_N \prod_{i=1}^p \prod_{\substack{v \subseteq V\backslash\{N\} \\ |v|=k_i-1 }}CZ_v \rangle\\
        &=\langle \underbrace{0...0}_\text{m}| \otimes \langle \underbrace{+...+}_\text{N-m-1}|  \prod_{i=1}^p \prod_{\substack{v \subseteq V\backslash\{N\} \\ |v|=k_i-1 }}CZ_v |\underbrace{0...0}_\text{m}\rangle \otimes| \underbrace{+...+}_\text{N-m-1}\ \rangle\\
        &=\langle \underbrace{+...+}_\text{N-m-1}|  \prod_{i=1}^p \prod_{\substack{v \subseteq V\backslash\{1,...,m, N\} \\ |v|=k_i-1 }}CZ_v | \underbrace{+...+}_\text{N-m-1}\ \rangle\\  
        &= \frac{1}{\sqrt{2^{N-m-1}}} \sum_{\tau \in \{0,1\}^{N-m-1}} \langle \tau  | \frac{1}{\sqrt{2^{N-m-1}}} \sum_{\tau' \in \{0,1\}^{N-m-1}} (-1)^{n_{\mathbf{k}-1}(\tau')} | \tau' \rangle\\
        &= \frac{1}{2^{N-m-1}} \sum_{\tau \in \{0,1\}^{N-m-1}} (-1)^{n_{\mathbf{k}-1}(\tau)} 
    \end{split}
    \label{eq base}
\end{equation}
\end{widetext}
where $CZ_e\big( |0_{e'}\rangle \otimes |\tau'\rangle \big) = |0_{e'}\rangle \otimes CZ_{e\backslash e'}|\tau'\rangle$ is utilized to derive the third line, and \eqref{computational basis hypergraph} to the fourth line. Comparing the two expectation-value expressions \eqref{eq base} and \eqref{eq coef} leads to

\begin{equation}
   \frac{1}{2^{N-m-1}} \sum_{\tau \in \{0,1\}^{N-m-1}} (-1)^{n_{k-1}(\tau)}=  C_0 + \sum_{n=1}^{m}C_n {m \choose n}
    \label{eq master}.
\end{equation}

To derive an explicit expression for \( C_n \), it is convenient to introduce the function
\begin{equation*}
    f_{\mathbf{k}}(m) = \frac{1}{2^{N - m - 1}} \sum_{\tau \in \{0,1\}^{N - m - 1}} (-1)^{n_{\mathbf{k}-1}(\tau)}.
\end{equation*}
This function can be rewritten as
\begin{equation}
    f_{\mathbf{k}}(m) = \frac{1}{2^{N - 1 - m}} \sum_{s = 0}^{N - 1 - m} \binom{N - 1 - m}{s} (-1)^{\sum_i\binom{s}{k_i - 1}},
    \label{f optimized}
\end{equation}
because for each \( s \in \{0, \dots, N - 1 - m\} \), there are \( \binom{N - 1 - m}{s} \) distinct strings \( \tau \in \{0,1\}^{N - 1 - m} \) with Hamming weight \( |J(\tau)| = s \). This correspondence yields a more efficient numerical procedure to evaluate \( f_{\mathbf{k}}(m) \).

Notice that $C_0=f_{\mathbf{k}}(0)$, $C_1=f_{\mathbf{k}}(1)-f_{\mathbf{k}}(0)$ and $C_2=f_{\mathbf{k}}(2)-2f_{\mathbf{k}}(1)+f_{\mathbf{k}}(0)$. Assuming that there is a positive integer $m$ such that
\begin{equation}
     C_n= \sum_{r=0}^n(-1)^{n-r} {n \choose r}f_{\mathbf{k}}(r).
    \label{eq final}
\end{equation}
for every $n\leq m$, then using \eqref{eq master}
\begin{equation}
\begin{split}
    C_{m+1} =& f_{\mathbf{k}}(m+1)- \sum_{n=0}^{m}{m+1 \choose n}C_n\\
    % =& f_{\mathbf{k}}(m+1)- \sum_{n=0}^{m} {m+1 \choose n} \sum_{r=0}^n (-1)^{n-r} {n \choose r}f_{\mathbf{k}}(r) \\
    =& f_{\mathbf{k}}(m+1)- \sum_{r=0}^{m}\sum_{n=r}^m (-1)^{n-r}{m+1 \choose n} {n \choose r}f_{\mathbf{k}}(r)
\end{split}
    \label{checagem}
\end{equation}

Now consider the term of the sum in \eqref{checagem} for $r$ fixed, using the expression 
\begin{equation*}
   \sum_{n=r}^m (-1)^{n}{m+1 \choose n} {n \choose r}
    =  (-1)^{m+1}{m+1 \choose r} 
\end{equation*}
proven at the appendix \ref{app: Binomial} and substituting in \eqref{checagem} leads to
\begin{equation*}
    \begin{split}
        C_{m+1}&= f_{\mathbf{k}}(m+1)-\sum_{r=0}^m (-1)^{-r} (-1)^m{ m+1 \choose r}f_{\mathbf{k}}(r)\\
        &= \sum_{r=0}^{m+1} (-1)^{m+1-r}{ m+1 \choose r}f_{\mathbf{k}}(r)
    \end{split}
\end{equation*}
therefore, by induction, \eqref{eq final} is valid for any $m\leq n\leq N-1$.

Our goal was to construct a Bell inequality for hypergraph states based on the coefficients obtained from their local operator representation, generalizing the method of \cite{Baccari_2020}. This approach allows the definition of a Bell functional that corresponds to a sum of stabilizers of the hypergraph for a suitable choice of measurement strategy. However, for all tested functionals, the HS did not achieve the maximal quantum violation; in fact, they failed to violate the local bound at all. This behavior is attributed to the presence of negative coefficients \( C_m \).

Numerical analysis confirmed that for all tested cases, at least one coefficient becomes negative whenever \( k > 2 \) (i.e., when the system cannot be reduced to an ordinary graph). 
An interesting case arises when \( C_0 = 0 \), as this coefficient corresponds to a marginal term in the Bell functional. The quantum bound of a Bell functional typically exhibits a non-linear dependence on the coefficient of such a marginal term~\cite{tiltedCHSH}. Therefore, it can be convenient to restrict the analysis to cases where \( C_0 = 0 \). This condition holds when \( k = 2^n + 1 \) and \( N = 2^{n+1} m \), for any positive integers \( n \) and \( m \) for a $k-$uniform complete HS. (see Appendix~\ref{app: proof C_0=0}). 

\section{Conclusions}

%In this work, we presented a novel approach to writing the generalized controlled gate in terms of local operators and expressing the nonlocal stabilizer formalism in a manner that applies to uniform complete HS. Such an alternative approach can be useful for self-testing these states and writing Bell operators for generalized GHZ-like states, such as the symmetric hypergraphs analyzed in this paper. Applications on quantum error correction and stabilizer codes using this approach might also be appealing for further investigation.

In this work, we have successfully developed a novel method to express the intrinsically nonlocal stabilizers of $k$-uniform complete hypergraph states as a linear combination of local operators. Our formulation provides an explicit, closed-form expression  for the expansion coefficients $C_m$, which depend on the uniformity $k$ and the number of vertices $N$. This was achieved by first deriving a local decomposition of the generalized $CZ$ gate  and then applying it to the full stabilizer $g_l$, leveraging the symmetries of complete hypergraphs. We provided an explicit example for the $N=4, k=3$ hypergraph state, demonstrating the practical application of our formalism.

A primary motivation for this investigation was to construct new Bell inequalities for hypergraph states by generalizing the successful method used for graph states in \cite{Baccari_2020}. This approach relies on defining a Bell functional from the stabilizer, for which our local expansion  is perfectly suited.

However, our analysis revealed a significant obstacle. When we tested the resulting Bell functionals, the hypergraph states not only failed to reach the maximal quantum violation, but in fact failed to violate the local bound at all. We identified the root cause of this failure: the presence of negative coefficients $C_m$ for all tested cases where $k>2$. This negative sign structure hinders the direct application of the method from \cite{Baccari_2020}, which implicitly relies on positive coefficients. This finding is a key part of our discussion, as it demonstrates that a simple generalization from graphs ($k=2$) to hypergraphs ($k>2$) is not viable for this specific purpose. As a result, the formulation of Bell inequalities in this case requires an alternative approach.

Despite this specific application not yielding a direct violation, the local operator expansion we derived remains a valuable and fundamental tool. The stabilizer formalism is central to many areas of quantum information beyond nonlocality. Our method may, as we initially hoped, find applications in quantum error correction. The ability to express nonlocal checks in terms of local operators is a powerful concept, and our formalism could offer new pathways for designing or analyzing stabilizer codes based on these complex states.

Furthermore, this work opens several avenues for future research.    While the method from \cite{Baccari_2020} failed, our local expansion could be the starting point for a different type of Bell functional. The explicit local form of the stabilizers may prove useful for self-testing these states, which is a key protocol for device-independent certification. Moreover, a promising direction is to investigate the special cases we identified where the marginal term $C_0=0$. Since the properties of Bell functionals can depend non-linearly on this term , these specific hypergraphs might be a more convenient starting point for analysis.

While the direct path to constructing Bell inequalities from local stabilizer expansions appears to be blocked by the sign structure of the coefficients, our work provides a novel and complete description of $k$-uniform complete hypergraph state stabilizers in terms of local operators. This fundamental description offers new insights for exploring these states within the stabilizer formalism and opens the door for new applications in quantum error correction and other areas of quantum information.

\section*{Acknowledgements}

G. Moniz acknowledges financial support from the  Fundação de Amparo à Pesquisa do Estado de São Paulo (FAPESP), BCO Doutorado Direto, through Grant No. 2024/09165-7.

V. Salem acknowledges the Spanish MCIN with funding from the European Union Next Generation EU (PRTRC17.I1) and Consejería de Educación from Junta de Castilla y León through the QCAYLE project and grant No. PID2023-148409NS-i00 MTM funded by AEI/10.13039/501100011033, and RED2022-134301-T.

B. Amaral acknowledges financial support from the Instituto Serrapilheira, Chamada No. 4 2020. and the  Fundação de Amparo à Pesquisa do Estado de São Paulo, Auxílio à Pesquisa—Jovem Pesquisador, through Grant No. 2020/06454-7 (until September 2025).

%%%References
\bibliographystyle{apsrev4-2}
\bibliography{references.bib}

%%%Appendix
\appendix
\onecolumngrid

\section{Conditions for $C_0=0$}
\label{app: proof C_0=0}

Since $C_0=f_{\mathbf{k}}(0)$ due to \eqref{eq master} and using the expression for $f_{\mathbf{k}}$ given at \eqref{f optimized}, we have that
\begin{equation}
    C_0=\frac{1}{2^{N-1}}\sum_{s=0}^{N-1} {N-1\choose s} (-1)^{s \choose k-1}=0
    \label{C_0=0}
\end{equation}
implies that $N$ is even and
\begin{equation}
    {r-1 \choose k-1} \neq {N-r \choose k-1},\ mod(2)
    \label{eq modulo}
\end{equation}
for every $1 \leq r \leq N$, since to satisfy \eqref{C_0=0} for each positive term there must be another negative term with the same coefficient and 
\begin{equation*}
    {N-1\choose r-1}={N-1\choose N-r}.
\end{equation*}

First consider the case where $1\leq r \leq 2^n$, reordering the denominator as $$2^n!= r(r+1)...(2^n)1 ...(r-1)$$ the binomial expression can be written as 
\begin{equation*}
    \begin{split}
        {2^{n+1}-r \choose 2^n}&=\largep{\frac{2^{n+1}-r}{r}}...\largep{\frac{2^{n+1}-2^n}{2^n}}...\\&...\largep{\frac{2^{n+1}-2^n-r+1}{r-1}} \\
        &= \prod_{p=r}^{2^n} \frac{2^{n+1}-p}{p}\prod_{j=1}^{r-1} \frac{2^{n}-j}{j}
        \label{expansion binomial}
    \end{split}
\end{equation*}
evidentiating that the binomial is odd.
While 
\begin{equation*}
        {r-1 \choose 2^n} = 0
\end{equation*}
which is even.\\

Observe that it suffices to analyze $r,\ mod(2^{n+1})$, because if $r \equiv l, \ mod(2^{n+1})$ then exist $m'$ non-negative integer such that $r=2^{n+1}m'+l$, ($m'<m$ since $r\leq 2^{n+1}m$, define $d=m-m'>0$) and taking $1\leq l \leq 2^n$, it is possible to follow analogously as in \eqref{expansion binomial}:
\begin{equation}
    \begin{split}
        {2^{n+1}m-r \choose 2^n}
        &= {2^{n+1}d -l
        \choose 2^n} \\
        &= \prod_{p=l}^{2^n} \frac{2^{n+1}d-p}{p}\prod_{j=1}^{r-1} \frac{2^{n}d-j}{j}
    \end{split}
\end{equation}
which again is odd.\\

While it is even for $r \equiv l, \ mod(2^{n+1})$, when $2^n +1 \leq l \leq 2^{n+1}$. Taking $l= 2^n + y$
and the same definitions as before, ordering $2^n!$ starting with $y$

\begin{equation}
\begin{split}
{2^{n+1}m-r \choose 2^n}
   &= {2^{n+1}m-(2^{n+1}m' + l) \choose 2^n} \\
   &= {2^{n+1}d - 2^n - y \choose 2^n} \\
   &= \frac{(2^{n+1}d - 2^n - y)(2^{n+1}d - 2^n - y - 1)\cdots(2^{n+1}d - 2^{n+1})}%
           {y(y+1)\cdots(2^n)} \\
   &\quad\times
      \frac{(2^{n+1}d - 2^{n+1} - 1)\cdots(2^{n+1}d - 2^{n+1} - y + 1)}%
           {(1)\cdots(y-1)} \\
    &= \prod_{p=y}^{2^n} 
      \frac{2^{n+1}d - 2^n - p}{p}\prod_{j=1}^{y-1}\frac{2^{n+1}d - 2^{n+1} - j}{j}
\end{split}
\end{equation}

which is even due to the term 
$\frac{2^{n+1}d-2^{n+1}}{2^n}=2(d-1)$.\\

Finally, observe the following congruence equivalences:
\begin{itemize}
    \item $r \equiv l, \ mod(2^{n+1})$, for $1\leq l \leq 2^n$ is equivalent to $r \equiv -l, \ mod(2^{n+1})$ for $2^n\leq l \leq 2^{n+1}-1$, then in this case 
    \begin{equation*}
        {r-1 \choose k-1} 
    \end{equation*} is even
    \item $r \equiv l, \ mod(2^{n+1})$, for $2^n +1 \leq l \leq 2^{n+1}$ is equivalent to $r \equiv -l, \ mod(2^{n+1})$, for $0 \leq l \leq 2^{n}-1$, then in this case 
    \begin{equation*}
        {r-1 \choose k-1} 
    \end{equation*} is odd
\end{itemize}
this verifies that equation \eqref{eq modulo} is satisfied for every $1 \leq r \leq N$, if $k-1=2^n$ and $N=2^{n+1}m$. Actually, this represents the only possible case, since to satisfy the \eqref{eq modulo} for any $r$ is necessary a symmetric relationship that only exists in this case.

\section{Binomial Identity}
\label{app: Binomial}

We present the algebra used in this work for the binomial identity as follows:

\begin{equation}
\begin{split}
    \sum_{n=r}^m (-1)^{n}{m+1 \choose n} {n \choose r}
     &= \sum_{n=r}^m (-1)^{n}\frac{(m+1)!}{(m+1-n)!n!} \frac{n!}{(n-r)!r!} \\ 
    &= \sum_{n=r}^m (-1)^{n}{m+1 \choose r} \frac{(m+1-r)!}{(m+1-n)!(n-r)!} \\
    &= {m+1 \choose r} \sum_{n=r}^m (-1)^{n}{m+1-r \choose n-r} \\
    &= {m+1 \choose r} \sum_{l=0}^{m-r} (-1)^{r+l}{m-r+1 \choose l} \\
    &= (-1)^{r}{m+1 \choose r} \left[-(-1)^{m-r+1} + \sum_{l=0}^{m-r+1} (-1)^{l}{m-r+1 \choose l}\right] \\
    &= (-1)^{r}{m+1 \choose r} \left[-(-1)^{m-r+1} + (1-1)^{m-r+1}\right] \\
    &= -(-1)^m{m+1 \choose r}
\end{split}
\end{equation}

\begin{equation}
\begin{split}
    CZ_v =& \mathbb{I} - 2|1_{v_1}...1_{v_n} \rangle \langle1_{v_1}...1_{v_n}|\\
    =& \mathbb{I} - 2 \largep{\frac{I_{v_1}-Z_{v_1}}{2}}... \largep{\frac{\mathbb{I}_{v_n}-Z_{v_n}}{2}}\\
    =& \mathbb{I} - \frac{1}{2^{|v|-1}}\bigotimes_{i\in v}(\mathbb{I}_i - Z_i)\\
    =&  \mathbb{I} - \frac{1}{2^{|v|-1}}\largep{\mathbb{I} + \sum_{m=1}^{|v|}(-1)^m\largep{\sum_{\substack{v' \subset v \\ |v'|=m}}\bigotimes_{i\in v'}Z_i}}\\
    =& \largep{1 - \frac{1}{2^{|v|-1}}}\mathbb{I}+ \sum_{m=1}^{|v|}\frac{(-1)^{m+1}}{2^{|v|-1}}\largep{\sum_{\substack{v' \subset v \\ |v'|=m}}\bigotimes_{i\in v'}Z_i}
    \label{CZ expansion}
\end{split}
\end{equation}

\end{document}